\documentclass[aps,prb,showpacs,twocolumn]{revtex4}
\usepackage{graphicx}
\usepackage{bm}
\begin{document}
\title{Low temperature electron-phonon resonance in dc-current-biased
    two-dimensional electron systems}
\author{X. L. Lei}
\affiliation{Department of Physics, Shanghai Jiaotong University,
1954 Huashan Road, Shanghai 200030, China}

\begin{abstract}
Effects of resonant acoustic phonon scattering on magnetoresistivity are examined 
in two-dimensional electron systems at low temperatures by
using a  balance-equation magnetotransport scheme direct controlled by the current. 
The experimentally observed resonances in linear resistivity are shown to result from 
the conventional bulk phonon modes in a GaAs-based system, without invoking leaky interface phonons.
Due to quick heating of electrons, phonon resonances can be dramatically enhanced
by a finite bias current. When the electron drift velocity increases to the speed of sound, 
additional and prominent phonon resonance peaks begin to emerge.
As a result, remarkable resistance oscillation and negative differential resistivity can
appear in nonlinear transport in a modest mobility sample at low temperatures, which is in agreement with %%@
recent experiments.

\end{abstract}


\maketitle

\section{Introduction}

Low-temperature magnetoresistance oscillations related to linear and nonlinear transport of electrons in
high Landau levels of high-mobility two-dimensional systems, induced 
by microwave radiation\cite{Ryz,Mani,Zud03,Durst,Lei03,Dietel05,Torres05,Dmitriev05,Ina-prl05} or by direct %%@
current excitation,\cite{Yang2002,Bykov-prb05,WZ-prb07,JQZ-prb07,Lei-07-1,Vav-prb07,Ina-apl07,Kunold-07} have %%@
attracted a great deal of attention in the past few years. Despite the fact that detailed microscopic %%@
mechanisms are still
under debate, there is almost no objection to refer these oscillatory magnetoresistances 
mainly to impurity or disorder scatterings 
and the direct phonon contributions to resistivity are believed to be negligible 
in these systems at such low temperatures.

Recently, the magnetophonon resonance in semiconductors, previously known to result from electron coupling 
with optic phonons and can be observed only at high temperatures and high magnetic %%@
fields,\cite{Gurevich61,Tsui80}
has been demonstrated to occur at temperatures as low as $T\sim 3$\,K and lower magnetic fields 
in GaAs-based heterosystems.\cite{Zudov-prl-01,Ponom2001,Yang-physica-02,Bykov-JETP05}
The resonant magnetoresistance was detected and referred to as electron scatterings 
by two leaky interface phonon modes that have sound velocities 
$v_s\simeq 2.9$\,km/s and $v_s\simeq 4.4$\,km/s,\cite{Zudov-prl-01,Ponom2001,Yang-physica-02} or by a
single leaky interface phonon mode that has a velocity $v_s= 5.9$\,km/s.\cite{Bykov-JETP05}
Very recently, Zhang {\it et al\,}\cite{WZ-prl08} found that 
acoustic phonon-induced resistance resonances are dramatically
enhanced in the nonlinear dc response and a finite current can strongly modify the phonon resonance 
behavior, transforming 
resistance maxima into minima and back.\cite{WZ-prl08}
These phonon-related resistance oscillations remain poorly understood, especially
the exact resonant condition, relative contributions of different modes and how they are affected by 
temperature and current.

In this paper, we present a systematic  
analysis on nonlinear magnetotransport in GaAs-based semiconductors
with a microscopic balance-equation scheme directly controlled by the current,
which takes into account electron couplings with impurity, bulk longitudinal and transverse acoustic 
phonons as well as polar optic phonons. 
Due to the quick rise of electron temperature, the phonon resonances are dramatically enhanced
by a finite current. 
When electron drift velocity $v$ gets into the supersonic regime ($v\geq v_s$, the speed of sound),
additional magnetophonon resonance peaks emerge.
As a result, a remarkable resistance oscillation and a negative differential resistivity 
appear in the nonlinear magnetotransport in a modest mobility sample at low temperatures.

\section{Balance equations for nonlinear magnetotransport}

We consider a quasi-two-dimensional (2D) system consisting of $N_{\rm s}$ electrons in a unit area of 
an $x$-$y$ plane. These electrons, 
subjected to a uniform magnetic field ${\bm B}=(0,0,B)$ along the $z$ direction
and a uniform electric field ${\bm E}$ in the $x$-$y$ plane, are
scattered by random impurities and by phonons in the lattice.
In terms of the center-of-mass momentum and coordinate defined as 
${\bm P}\equiv\sum_j {\bm p}_{j\|}$ 
and ${\bm R}\equiv N_{\rm s}^{-1}\sum_j {\bm r}_{j}$,  
with ${\bm p}_{j\|}\equiv(p_{jx},p_{jy})$ and ${\bm r}_{j}\equiv (x_j,y_j)$
being the momentum and coordinate of the $j$th electron in the 2D plane,
and the relative electron momentum and coordinate 
${\bm p}_{j\|}'\equiv{\bm p}_{j\|}-{\bm P}/N_{\rm s}$ and 
${\bm r}_{j}'\equiv{\bm r}_{j}-{\bm R}$, 
the Hamiltonian $H$ of this coupled electron-phonon system can be written 
as the sum of a center-of-mass part $H_{\rm cm}$,
a relative electron part $H_{\rm er}$,\cite{Ting,Lei85,Lei851,Lei852}
\begin{eqnarray}
&&\hspace{-0.5cm}H_{\rm cm}=\frac 1{2N_{\rm s}m}\big({\bm P}-N_{\rm s}e{\bm A}({\bm
R})\big)^2-N_{\rm s}e{\bm E}\cdot {\bm R},\\
&&\hspace{-0.5cm}H_{\rm er}=\sum_{j}\left[\frac{1}{2m}\Big({\bm p}_{j\|}'-e{\bm A}
({\bm r}_{j}')\Big)^{2}+\frac{p_{jz}^2}{2m_z}+V(z_j)\right]\nonumber\\
&&\hspace{1.1cm}+\sum_{i<j}V_{\rm c}({\bm r}_{i}'-{\bm r}_{j}',z_i,z_j),\,\,\,\,\,\,\, \label{her}
\end{eqnarray}
and a phonon part $H_{\rm ph}=\sum_{{\bm q}\lambda}{\it \Omega}_{{\bm q}\lambda}
b_{{\bm q}\lambda}^{\dag}b_{{\bm q}\lambda}^{}$, 
together with electron-impurity and electron-phonon interactions as follows:
\begin{eqnarray}
\label{hei} H_{\rm ei}&=&\sum_{{\bm q}_{\|},a} u({\bm q}_{\|},z_a)\,{\rm e}^{-i{\bm q}_{\|}
\cdot{\bm r}_{\!a}}\,{\rm e}^{i{\bm q}_{\|}
\cdot{\bm R}}\rho_{{\bm q}_{\|}},\\
\label{hep} H_{\rm ep}&=&\sum_{{\bm q},\lambda} M({\bm q},\lambda)I(q_z)(b_{{\bm q}\lambda}^{}
+b_{-{\bm q}\lambda}^{\dag})
\,{\rm e}^{i{\bm q}_{\|}\cdot{\bm R}}\rho_{{\bm q}_{\|}}.
\end{eqnarray}
Here, ${\bm A}({\bm r})$, the in-plane component of the vector potential of the uniform magnetic field, 
is linear in the spatial coordinate ${\bm r}=(x,y)$;
$m$ and $m_z$ are, respectively, the electron effective mass parallel and perpendicular 
to the plane; $V(z)$ and $V_c({\bm r}_{i}'-{\bm r}_{j}',z_i,z_j)$ stand for the confined 
and Coulomb potentials, respectively 
$\rho_{{\bm q}_{\|}}=\sum_j {\rm e}^{i{\bm q}_{\|}\cdot{\bm r}'_{j{\|}}}$ 
is the density operator of the two-dimensionalo (2D) relative electrons; 
$u({\bm q}_{\|},z_a)$ is the effective potential of the $a$th impurity
located at $({\bm r}_{\!a},z_a)$ in the 2D Fourier space;
 $b_{{\bm q}\lambda}^{\dag}(b_{{\bm q}\lambda}^{})$ is the creation 
(annihilation) operator of the bulk phonon with wave vector 
${\bm q}=({\bm q}_{\|},q_z)=(q_x,q_y,q_z)$ in branch $\lambda$
that has an energy ${\it \Omega}_{{\bm q}\lambda}$; 
$M({\bm q},\lambda)$ is the matrix element of
the electron-phonon interaction in the three-dimensional (3D) plane-wave representation; 
and $I(q_z)$ is a form factor of the quasi-2D electron.\cite{Lei851}
Here, for simplicity, we have assumed that the quasi-2D electrons occupy only the lowest subband and
thus the subband summation indices in Eqs.\,(\ref{hei}) and (\ref{hep}) are neglected.\cite{Lei851}

The separation of the electron Hamiltonian into a center-of-mass part 
and a relative electron part amounts to looking at electrons 
in a reference frame moving with their center of mass.   
The most important feature of this separation is that a spatially uniform electric field 
shows up only in $H_{\rm cm}$, and that $H_{\rm er}$ is the Hamiltonian of a many particle system 
subject to a perpendicular magnetic field {\it without the electric field}.
This enables us to deal with relative electrons in the magnetic field
{\it without tilting the Landau levels}.  
The coupling between the center of mass and relative electrons is shown by  
the factor ${\rm e}^{i{\bm q}_{\|}\cdot{\bm R}}$ inside the momentum 
summation in $H_{\rm ei}$ and $H_{\rm ep}$.   
The moving center-of-mass assisted transitions of relative electrons between different Landau levels 
provide the major mechanism for the current-driven magnetotransport.
 
Our treatment starts with
the Heisenberg operator equations for the rate of change in the center-of-mass velocity 
${\bm V}=-i[{\bm R},H]$, and that for the relative electron energy $H_{\rm er}$ as follows:
\begin{eqnarray}
\dot{\bm V}&=&-i[{\bm V},H],\label{opv}\\
%\end{equation}
%and for the rate of change of the relative electron energy 
%\begin{equation}
\dot{H}_{\rm er}&=&-i[H_{\rm er},H].\label{oph}
\end{eqnarray} 
When the electron-impurity and electron-phonon couplings are weak in comparison with
the internal thermalization of relative electrons and that of phonons, it is good 
enough to carry out the statistical average of the above operator equations  
to leading orders in $H_{\rm ei}$ and $H_{\rm ep}$.
For this purpose, we only need to know the distribution of relative electrons 
and phonons without being perturbed by $H_{\rm ei}$ or $H_{\rm ep}$.
The distribution function of the relative electron system described 
by Hamiltonian (\ref{her})
without an electric field should be an isotropic Fermi-type function with a single
temperature $T_{\rm e}$. The phonon system, which is assumed to be in an equilibrium
state, has a Bose distribution with lattice temperature $T$.   
Such a statistical average of the above operator equations yields 
the following force and energy balance equations in the steady state, which has a constant
average drift velocity ${\bm v}$:
\begin{equation}
N_{\rm s}e{\bm E}+N_{\rm s} e ({\bm v} \times {\bm B})+
{\bm f}({\bm v})=0,\label{eqforce}\\
\end{equation}
\begin{equation}
{\bm v}\cdot {\bm f}({\bm v})+ w({\bm v})=0. \label{eqenergy}
\end{equation}
Here, ${\bm f}({\bm v})={\bm f}_{\rm i}({\bm v})+{\bm f}_{\rm p}({\bm v})$ 
is the frictional force experienced by the electron center of mass due to impurity and phonon 
scatterings, given by
\begin{eqnarray}
&&{\bm f}_{\,\rm i}({\bm v})=\sum_{{\bm q}_\|} {\bm q}_\| \left| U({\bm q}_\|)\right| ^{2}
{\it \Pi} _{2}({\bm q}_\|,\omega_0),\label{forcei}\\
%\end{equation}
%\begin{equation}
&&{\bm f}_{\rm p}({\bm v})=2\sum_{{\bm q},\lambda} {\bm q}_\| 
\left|{M}({{\bm q},\lambda})\right|^{2}|I(q_z)|^2
{\it \Pi}_{2}({\bm q}_{\|},{\it \Omega}_{{\bm q}\lambda}+\omega_0)\nonumber\\
&&\hspace{2cm}\times\left[n\!\left (\frac{{\it \Omega}_{{\bm q}\lambda}}{T}\right )
-n\!\left (\frac{{\it \Omega}_{{\bm
q}\lambda}+\omega_0}{T_{\rm e}} \right )\right],\,\,\,
 \label{forcep}
\end{eqnarray}
and $w({\bm v})$ is the electron energy-loss rate to the lattice due to electron-phonon interactions 
with an expression obtained from the right-hand side of Eq.\,(\ref{forcep}) 
by replacing the ${\bm q}_\|$ factor with ${\it \Omega}_{{\bm q}\lambda}$. 
In these equations, $\omega_0\equiv{\bm q}_\|\cdot {\bm v}$,
$|U({\bm q}_\|)|^2$  is the effective average impurity scattering potential, 
and $|{M}({\bm q},\lambda)|^2|I(q_z)|^2$ is the effective coupling matrix element 
between a $\lambda$-branch 3D phonon
and a quasi-2D electron, 
${\it \Pi}_2({\bm q}_\|,{\it \Omega})$ is the imaginary part of the 2D electron density 
correlation function 
at electron temperature $T_{\rm e}$ in the presence of the magnetic field, and
$n(x)\equiv 1/({\rm e}^x-1)$ is the Bose function.
The effect of interparticle Coulomb interaction is included in the density correlation function 
to the degree of electron level broadening and screening. 
With the screening statically considered in the effective impurity and phonon potentials,
the remaining ${\it \Pi}_2({\bm q}_{\|}, {\it \Omega})$ function in Eqs.\,(\ref{forcei}) and (\ref{forcep})
is that of a noninteracting  2D electron gas  
in the magnetic field, which can be written in the Landau representation as\cite{Ting}
\begin{eqnarray}
&&\hspace{-0.7cm}{\it \Pi}_2({\bm q}_{\|},{\it \Omega}) =  \frac 1{2\pi
l_{\rm B}^2}\sum_{n,n'}C_{n,n'}(l_{\rm B}^2q_{\|}^2/2) 
{\it \Pi}_2(n,n',{\it \Omega}),
\label{pi_2q}\\
&&\hspace{-0.7cm}{\it \Pi}_2(n,n',{\it \Omega})=-\frac2\pi \int d\varepsilon
\left [ f(\varepsilon )- f(\varepsilon +{\it \Omega})\right ]\nonumber\\
&&\,\hspace{2cm}\times\,\,{\rm Im}G_n(\varepsilon +{\it \Omega})\,{\rm Im}G_{n'}(\varepsilon ),
\label{pi_2ll}
\end{eqnarray}
where $l_{\rm B}=\sqrt{1/|eB|}$ is the magnetic length,
$
C_{n,n+l}(Y)\equiv n![(n+l)!]^{-1}Y^l{\rm e}^{-Y}[L_n^l(Y)]^2
$
with $L_n^l(Y)$ the associate Laguerre polynomial, $f(\varepsilon
)=\{\exp [(\varepsilon -\mu)/T_{\rm e}]+1\}^{-1}$ is the Fermi 
function at electron temperature $T_{\rm e}$, 
and ${\rm Im}G_n(\varepsilon )$ is the density-of-states of the broadened Landau level $n$.

We model the electron density-of-states function 
with a Gaussian-type form for both overlapped and separated 
Landau levels ($\varepsilon_n=n\omega_c$ is the center of the $n$th Landau level;
$\omega_c=eB/m$ is the cyclotron frequency) as follows:\cite{Ando82}
\begin{equation}
{\rm Im}G_n(\varepsilon)=-(2\pi)^{\frac{1}{2}}{\it \Gamma}^{-1}
\exp[-2(\varepsilon-\varepsilon_n)^2/{\it \Gamma}^2].
\label{Gauss}
\end{equation}
The half-width $\it \Gamma$, or, the life-time or the quantum scattering time,
 $\tau_s=1/2{\it \Gamma}$, of the Landau level, which should be 
determined by  electron-impurity, electron-phonon and electron-electron scatterings in the system,
is magnetic-field $B$ and temperature $T$ dependent. We treat it as a semiempirical parameter,
which will serve as the only adjustable parameter in the present investigation. 

At lattice temperature $T$, the energy balance [Eq.\,(\ref{eqenergy})] yields the electron temperature
$T_{\rm e}$ for a given carrier drift velocity ${\bm v}$ at a given magnetic field. 
Then, with this $T_{\rm e}$, the force balance [Eq.\,(\ref{eqforce})]
determines the relation between ${\bm v}$, ${\bm B}$ and ${\bm E}$, 
i.e., the longitudinal and transverse resistivities in the magnetotransport.

Such a formulation indicates that the
carrier drift velocity  ${\bm v}$
is the basic physical quantity that controls the nonlinear magnetotransport.
The frictional force ${\bm f}({\bm v})$ and energy dissipation rate $w({\bm v})$
solely depend on the drift velocity
${\bm v}$ at a given magnetic field $B$, 
while the electric field only plays a role in balancing the frictional force.
Thus, the resistivity is directly determined by the scattering mechanisms 
and by the drift velocity or the current density,
rather than by the electric field. 
Equations (\ref{eqforce}) and (\ref{eqenergy}) are conveniently applied 
to current-driven magnetotransport of any configuration, in which the current is 
an experimentally directly controlled quantity. 
For an isotropic system where the frictional force is in the opposite direction of 
the drift velocity ${\bm v}$
and the magnitudes of both the frictional force and the energy-loss rate depend only on 
$v\equiv |{\bm v}|$, we can write ${\bm f}({\bm v})=f(v){\bm v}/v$ and 
$w({\bm v})=w(v)$.  
In the Hall configuration with velocity ${\bm v}$ in the $x$ direction
[${\bm v}=(v,0,0)$] or the current density $J_x=J=N_{\rm s}ev$ and $J_y=0$,
Eq.\,(\ref{eqforce}) gives, after $T_{\rm e}$ [thus, the $f(v)$ function] 
determined from Eq.\,(\ref{eqenergy}), $vf(v)+w(v)=0$,
the transverse resistivity $R_{yx}=B/N_{\rm s}e$, and the longitudinal resistivity $R_{xx}$ and 
the longitudinal differential resistivity $r_{xx}$ as
\begin{eqnarray} 
&&R_{xx}= -f(v)/(N_{\rm s}^2e^2v), \label{eqrxx}\\
&&r_{xx}=-({\partial f(v)}/{\partial v})/(N_{\rm s}^2e^2). \label{eqdr}
\end{eqnarray}
Note that, in principle,  since the Landau-level broadening and electron temperature
are determined by the simultaneous existence of all of the scattering mechanisms,
contributions to the total resistivity from different scattering mechanisms
are not independent. Nevertheless, it is still useful to formally write the total resistivity 
as a direct sum of separate scattering contributions by the respective components 
of the frictional force ${\bm f}({\bm v})$.

\section{Magnetophonon resonance in linear magnetoresistance}

In the numerical analysis we first concentrate on a GaAs-based heterosystem
with carrier sheet density $N_{\rm s}=4.8\times 10^{15}$m$^{-2}$ and zero-temperature
linear mobility $\mu_0=440$\,m$^2$/V\,s in the absence of a magnetic field, considering electron 
scatterings from bulk longitudinal acoustic (LA)  phonons (one branch, via the deformation 
potential and piezoelectric couplings with electrons) and transverse acoustic (TA) 
 phonons (two branches, via the piezoelectric coupling with electrons), 
 as well as from remote and background impurities. 
The coupling matrix elements are taken to be well known expressions\cite{Lei851}
with typical material parameters in bulk GaAs: 
electron effective mass $m=0.067\,m_{\rm e}$ ($m_{\rm e}$ is the
free electron mass),
longitudinal sound velocity $v_{\rm sl}=5.29\times 10^3$\,m/s, 
 transverse sound velocity $v_{\rm st}=2.48\times 10^3$\,m/s, acoustic 
deformation potential ${\it \Xi}=8.5$\,eV, piezoelectric constant $e_{14}=
1.41\times 10^9$\,V/m, dielectric constant $\kappa=12.9$, and
material mass density $d=5.31$\,g/cm$^3$. 
We take a magnetic-field-dependent ($B^{1/2}$) Landau-level half-width as follows:
\begin{equation}
{\it \Gamma}=\left[{8\alpha e\omega_c}/{\pi m \mu_0(T)}\right]^{1/2} \label{gam}
\end{equation}
 which is expressed
in terms of $\mu_{0}(T)$, the total linear mobility at lattice temperature $T$ 
in the absence of the magnetic field, and a broadening parameter $\alpha$ 
to take into account the difference in the transport scattering time 
from the broadening-related quantum lifetime.\cite{Mani,Durst}
A broadening parameter 
$\alpha=1.5$ is used in the calculation in Sects.\,III and IV, which 
corresponds to taking  Landau-level half-widths 
${\it \Gamma}=1.8$ and 2.4\,K, respectively, at lattice temperatures $T=5$ and 10\,K 
for magnetic field $B=0.5$\,T.

\begin{figure}
\includegraphics [width=0.40\textwidth,clip=on] {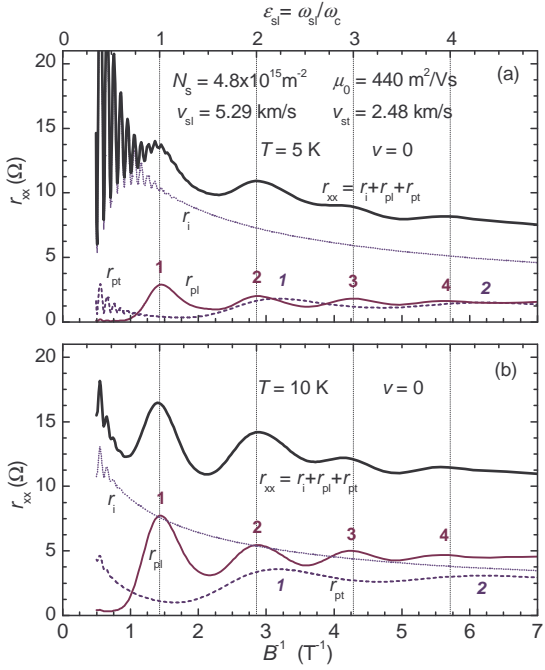}
\vspace*{-0.2cm}
\caption{(Color online) Linear ($v=0$) resistivities ($r_{xx}$, 
$r_{\rm i}$,  $r_{\rm pl}$, and $r_{\rm pt}$) at (a) $T=5$\,K (a)  and (b) $T=10$\,K
for a GaAs-based heterosystem with carrier sheet density 
$N_{\rm s}=4.8\times 10^{15}$m$^{-2}$ and zero-temperature
linear mobility $\mu_0=440$\,m$^2$/V\,s. The longitudinal and transverse 
sound velocities are taken to be  $v_{\rm sl}=5.29\times 10^3$\,m/s 
and $v_{\rm st}=2.48\times 10^3$\,m/s.}
\label{fig1}
\end{figure}

The calculated total linear ($v\rightarrow 0$) magnetoresistivity (differential resistivity)
$R_{xx}=r_{xx}$ is shown versus $1/B$ at lattice temperatures $T=5$\,K  and $T=10$\,K 
in Figs.\,1(a) and 1(b), 
together with separated contributions $r_{\rm i}$ from impurities,  $r_{\rm pl}$ from LA phonons, 
and $r_{\rm pt}$ from TA phonons: $r_{xx}=r_{\rm i}+r_{\rm pl}+r_{\rm pt}$.
The resistivity resonances clearly appear in $r_{\rm pl}$ and $r_{\rm pt}$ at both temperatures. 
This feature of magnetoresistivity stems from the property of 2D electron density correlation function
${\it \Pi}_2({\bm q}_\|, {\it \Omega}_{{\bm q}\lambda})$. In the case of low temperature 
($T_{\rm e}$ much less than the Fermi energy $\varepsilon_{\rm F}$)  
and large Landau-level filling factor ($\nu=\varepsilon_{\rm F}/\omega_c\gg 1$), 
${\it \Pi}_2({\bm q}_\|, {\it \Omega})$ is a periodic
function with respect to its frequency variable:  
${\it \Pi}_2({\bm q}_\|, {\it \Omega}+l\omega_c)={\it \Pi}_2({\bm q}_\|, {\it \Omega})$ 
for any integer $l$ of $|l|\ll \nu$.
On the other hand, under the same conditions (low temperature $T_{\rm e}\ll \varepsilon_{\rm F}$ 
and large filling factor $\nu\gg 1$), the major contributions to the summation in Eq.\,(\ref{pi_2q})
come from terms $n\sim n'\sim \nu$, and then the function $C_{n,n'}(x)$ has a sharp principal maximum
near $x=4\nu$.\cite{Scher66,JZhang-04} Therefore, as a function of the in-plane momentum $q_{\|}$,
 ${\it \Pi}_2({\bm q}_{\|},{\it \Omega})$ sharply peaks at
around $q_{\|} \simeq 2k_{\rm F}$. In view of the existence of the form factor $|I(q_z)|^2$,
which is  related to the wave function of the quasi-2D electron, only very small $q_z$ ($\ll k_{\rm F}$) 
can contribute to the integration in Eq.\,(\ref{forcep}), 
and thus, the contribution of the wave vector integration to
the phonon-induced frictional force  heavily weighs around $q  \simeq 2k_{\rm F}$.   
As a result of this phase-space weight distribution and the 3D phonon dispersion 
[${\it \Omega}_{{\bm q}\lambda}=v_{\rm s\lambda}q \, (\lambda= {\rm l, t})$], 
when $\omega_{\rm s\lambda}\equiv 2k_{\rm F}v_{\rm s\lambda}=l\omega_c$\,\,($l=1,2,3,...$),
a quasi-2D relative electron in any Landau level can be resonantly scattered by absorbing 
or emitting a phonon 
and jumps across $l$ Landau levels. The linear resistivity maxima then show up at 
\begin{equation}
\varepsilon_{\rm s\lambda}\equiv\omega_{\rm s\lambda}/\omega_c=l=1,2,3,... \label{sn}
\end{equation}
This is exactly what is seen in Figs.\,1(a) and 1(b), as well as in the lower part of Fig.\,2(a), 
where the $r_{\rm pl}$ maxima located at 
$\varepsilon_{\rm sl}=1,2,3,4$ and $r_{\rm pt}$ maxima at 
$\varepsilon_{\rm st}=1,2$ are labeled. The total resistivity $r_{xx}$ peaks at around
$\varepsilon_{\rm sl}=1,2,3,4$, which is essentially determined by the resonant scattering 
of LA phonons.
These results are in good agreement with experiments,\cite{Bykov-JETP05,WZ-prl08}
which indicates that the observed low-temperature magnetophonon resonances in linear 
magnetoresistance are
well explained by the ordinary single bulk LA phonon mode in GaAs with no
need to invoke leaky interface modes. 

\section{Magnetophonon resonance in nonlinear transport}

A finite current density $J$ or a finite drift velocity $v$ in the $x$ direction has two major effects.
First, it results in the electron heating and thus raises the 
rate of phonon emission. As a result, the phonon contributed resistivity and the oscillatory amplitude
of magneto-phonon resonance is enhanced with increasing bias current density.
The electron temperature is determined by the frictional force $f(v)$ and the energy-dissipation rate
$w(v)$ through the energy balance [Ea.\,(\ref{eqenergy})]. Generally, longitudinal acoustic phonons give 
the dominant contribution to 
$w(v)$ in GaAs-based systems at lattice temperatures considered in this paper ($T \leq 10$\,K), 
and the polar optic (LO) phonons are usually frozen. However, 
electron scattering from LO phonons should still be taken into account 
when the bias current density becomes strong that electron temperature $T_{\rm e}$ rises up to 
the order of $20$\,K, at which a weak emission of LO phonons can take place. 
These emitted LO phonons, although giving
little contribution to the resistivity itself, provide an additional efficient
energy dissipation channel to prevent the continuing rise in electron temperature.\cite{Lei-prb-05} 
Therefore, we take the LO-phonon scattering (via the Fr\"{o}hlich coupling electrons,
with an optical dielectric constant $\kappa_{\infty}=10.5$)
into account in the numerical calculation in nonlinear transport.
The calculated electron temperatures $T_{\rm e}$ in the case of $T=10$\,K  
at different bias drift velocities 
$2v/v_{\rm F}=0.001,0.002,0.003,0.004,0.005$, and 0.006 ($v_{\rm F}$ is the Fermi velocity 
of the 2D electron system), which correspond to current densities
$J=0.115,0.231,0.346,0.462,0.577$ and 0.693\,A/m, are shown in Fig.\,2(b) 
versus the magnetic field $B$ 
for the GaAs heterosystem introduced above.

\begin{figure}
\includegraphics [width=0.42\textwidth,clip=on] {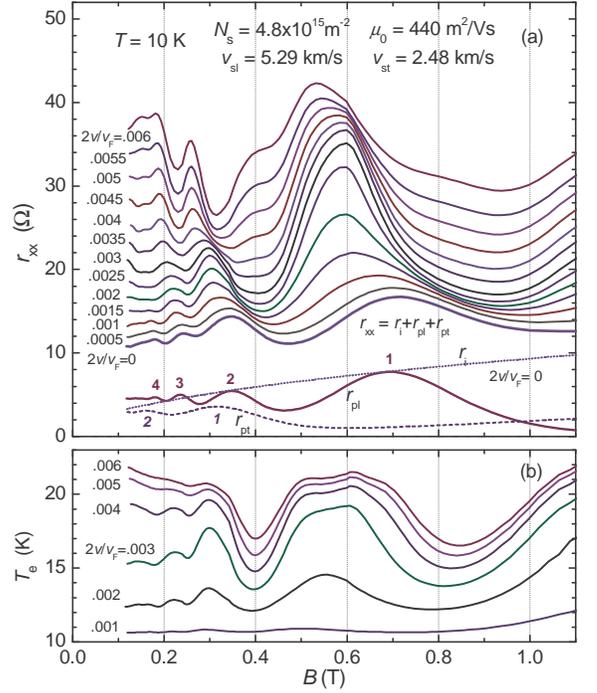}
\vspace*{-0.2cm}
\caption{(Color online) (a)  Differential resistivities $r_{xx}$  
and (b) electron temperature $T_{\rm e}$ are shown as functions of the magnetic-field strength $B$ 
at lattice temperature $T=10$\,K under different values of the bias drift velocity $v/v_{\rm F}$.
Here, $v_{\rm F}$ is the Fermi velocity of the 2D electron system, and $r_{xx}$ curves of
different $2v/v_{\rm F}$ values are vertically offset for clarity.}
\label{fig2}
\end{figure}

We see that, in addition to the concordant rise in electron temperature with increasing current density, 
at each bias drift velocity, $T_{\rm e}$ also exhibits resonance when changing magnetic field. 
The oscillatory peak-valley structure of $T_{\rm e}$
 periodically shows up, which has a period $\Delta (\omega_{\rm sl}/\omega_c)\approx 1$, 
indicating that it is also a magneto-phonon resonance mainly due to LA phonons. 
This $T_{\rm e}$ resonance stems 
from the periodicity of $f(v)$ and $w(v)$ functions in the energy-balance equation (\ref{eqenergy}).
Note that when the bias current density grows the oscillatory amplitude of $T_{\rm e}$ 
generally increases,
but the positions of $T_{\rm e}$ peaks (valleys) remain essentially the same in the $B$ axis. 
By comparing with the magnetophonon resonant linear resistivity  $r_{xx}$ 
[shown on the lower part of Fig.\,2(a)], the oscillatory structure of $T_{\rm e}$ peaks (valleys) appears 
roughly $\pi/2$ phase shift down along the $B$ axis. 

The differential resistivity $r_{xx}$ of the system at a given bias drift velocity $v$ is
obtained from the $\partial f(v)/\partial v$ function through Eq.\,(\ref{eqdr}),
with the electron temperature $T_{\rm e}$ determined above.
Since the peak (valley) positions of $T_{\rm e}$ are essentially fixed in the $B$ axis,
the variation of the resistivity maxima (minima) with changing bias velocity $v$
is mainly determined by its direct effect in the $\partial f(v)/\partial v$ function, 
as reflected in the frequency
 shift $\omega_0\equiv {\bm q}_{\|}\cdot {\bm v}$ in the argument 
 of the electron density correlation function. Physically,
because an extra energy ${\bm q}_\|\cdot{\bm v}$ is provided by 
the moving center of mass to the relative electrons during the scattering process, 
the transition rate of 
an electron from Landau level $n$ to $n'$ ($n'$ can be equal to or not equal to $n$) 
experiences a change due to impurity and phonon scatterings, which shows up   
through the $-{\bm q}_\|\cdot {\bm v}$ shift in, for instance, the 
${\it \Pi}_2({\bm q}_\|,{\it \Omega}_{{\bm q}\lambda}-{\bm q}_\|\cdot {\bm v})$
function in $-{\bm f}_{\rm p}$ (i.e., the resistivity).
The effect of such an energy shift in the impurity scattering case 
has been shown to induce an oscillatory differential magnetoresistance $r_{\rm i}$ 
that is controlled by the following parameter:\cite{Yang2002,WZ-prb07,Lei-07-1}
\begin{equation}
\varepsilon_{j}\equiv\omega_{j}/\omega_c,\,\,\,\,\omega_j\equiv 2k_{\rm F}v=\sqrt{8\pi/N_{\rm s}}J/e,
\end{equation}
which has peak positions at around $\varepsilon_{j}\approx 0,1,2,...$. 
In the case of acoustic phonon scatterings,
the resistivity maxima are expected to occur near the possible integer values of parameter 
$\varepsilon_{\rm s\lambda}-\varepsilon_{j}$, i.e.,
\begin{equation}
\varepsilon_{\rm s\lambda}-\varepsilon_{j}\approx l=0,\pm 1, \pm 2,... \label{sj}
\end{equation}
and this integer can be used as an identification for each magnetophonon resonance peak 
of the differential resistivity in nonlinear transport.
When drift velocity $v$ is smaller than the sound speed 
$v_{\rm s\lambda}$ ($\varepsilon_{j}<\varepsilon_{\rm s\lambda}$), 
the resonance condition (\ref{sj}) can be satisfied only for positive integers $l=1,2,3,4,...$, 
whence
the effect of a small current $J$ is roughly to change the resistivity maxima from  
$\varepsilon_{\rm s\lambda}\approx l$ in the linear case to 
\begin{equation}
\varepsilon_{\rm s\lambda}\approx l+\varepsilon_{j}.
\end{equation}
This means that for a given index number $l$, the peak position of magnetophonon resonance 
moves towards lower $B$ with increasing current density,
and for a given $J$ the shifts of peak positions in the $\varepsilon_{\rm s\lambda}$ axis 
are larger for larger index $l$ (lower $B$ field).
These features are clearly seen in Fig.\,2(a), where we show the calculated differential
resistivity $r_{xx}$ as a function of the magnetic field at $T=10$\,K
under different bias drift velocities
from $2v/v_{\rm F}=0$ to 0.006 in steps of 0.0005, corresponding to current densities
$J=0- 0.693$\,A/m in steps of 0.058\,A/m. 
We see that though within certain $v$ and $B$-field ranges, 
e.g. $2v/v_{\rm F}= 0.003 \sim .005$ and $B$ is around $6$\,T,
the peak positions of the resistivity may be somewhat influenced 
by the rapid change in the electron temperature because of the enhanced phonon contributions 
at higher $T_{\rm e}$;
the main trend of the resistivity peak shift with increasing $J$ remains.
The progress and movement of respective peaks compare favorably with recent 
experimental observation.\cite{WZ-prl08}

\begin{figure}
\includegraphics [width=0.45\textwidth,clip=on] {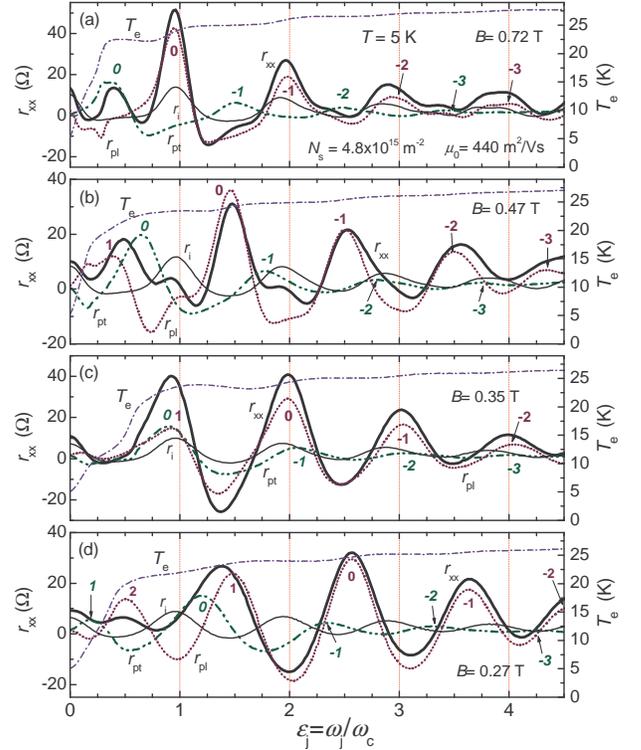}
\vspace{-0.2cm}
\caption{(Color online)  Differential resistivities ($r_{\rm i}$, $r_{\rm pl}$, $r_{\rm pt}$, and $r_{xx}$)
and electron temperature $T_{\rm e}$ are shown as functions of $\varepsilon_{j}$ 
at fixed magnetic fields $B=0.27,0.35, 0.47$, and 0.72\,T, respectively, at $T=5$\,K.
The number near the peak of  $r_{\rm pl}$ or $r_{\rm pt}$ is the value of the integer $l$ 
in the resonance condition (\ref{sj}) of the corresponding differential resistivity peak.}
\label{fig3}
\end{figure}

When $v$ becomes equal to or greater than the sound speed $v_{\rm s\lambda}$, 
the condition (\ref{sj}) can be satisfied by $l=0$ and negative integers, which indicates the occurrence 
of additional magnetophonon resonance peaks under a strong dc excitation.
Figure 3 presents the calculated differential resistivity $r_{xx}$ and 
electron temperature $T_{\rm e}$ at the lattice temperature $T=5$\,K
as functions of the current density 
in terms of $\varepsilon_{j}\equiv\omega_j/\omega_c$  at magnetic fields 
$B=0.27,0.35, 0.47$, and 0.72\,T. 
Due to the rapid rise in electron temperature $T_{\rm e}$ in this $\mu_0=440$\,m$^{2}$/V\,s 
system, acoustic phonon contribution to 
resistivity becomes significant when $\varepsilon_{j}>0.1$.
The prominent oscillations show up in resistivities $r_{\rm i}$, $r_{\rm pl}$, $r_{\rm pt}$, and $r_{xx}$,
all exhibiting a main oscillation period $\Delta \varepsilon_{j}\sim 1$.
However, since the peak positions of LA and TA phonon resistivities $r_{\rm pl}$ and 
$r_{\rm pt}$ in the $\varepsilon_{j}$ axis 
change with changing magnetic field, as indicated by condition (\ref{sj})
($\varepsilon_{j}\approx\varepsilon_{\rm s\lambda}-l$), 
the oscillating behavior of total resistivity $r_{xx}$ is strongly  
$B$-field dependent.
For instance, in the case of $B=0.35$\,T, the LA phonon resonance peaks 1,0,-1, and -2 
are essentially in phase with the TA phonon resonance peaks 0,-1,-2 and -3, and  close to the maxima
of the impurity resistivity $r_{\rm i}$ near $\varepsilon_{j}=1,2,3,4$, which lead to an
enhanced $r_{xx}$ with maxima around these positions [Fig.\,3(c)].
In the case of $B=0.47$\,T, where the LA resonance peaks 1,0,-1, and -2 are around 
$\varepsilon_{j}=0.5,1.5,2.5$, and 3.5
and the TA resonance peaks 0,-1,-2, and -3 are around $\varepsilon_{j}=0.7,1.7,2.7$, and 3.7,
the resulting $r_{xx}$ oscillation is out-of-phase with that of $r_{\rm i}$ oscillation 
and has secondary extremes [Fig.\,3(b)].
As a result, at fixed $\varepsilon_{j}$, e.g., $\varepsilon_{j}=2$, 
$r_{xx}$ exhibits maxima for $B=0.35$ and 0.72\,T while minima for $B=0.27$ and 0.47\,T.
Note that the resonant magnetophonon resistivity
at $v=v_{s\lambda}$ (the $l=0$ peak) is generally the largest among all of the resistivity
maxima in each type of phonon scattering, as can be seen from all four cases in Fig.\,3. 
It comes from phonon-induced intra-Landau-level 
scatterings of electrons, which
are allowed for all of the wavevectors, with the energy provided by the center of mass 
that has velocity  $v=v_{s\lambda}$.

\begin{figure}
\includegraphics [width=0.40\textwidth,clip=on] {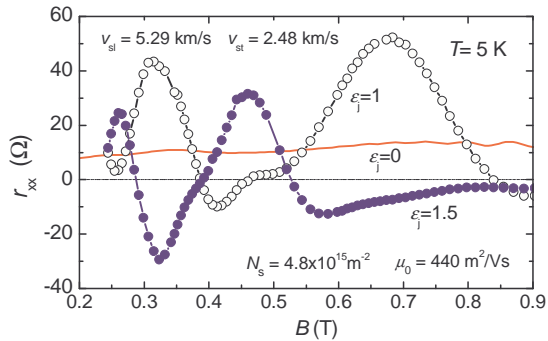}
\caption{(Color online)  Differential resistivity $r_{xx}$ is plotted versus magnetic field $B$ 
for fixed values of $\varepsilon_j=0$, 1, and 1.5 at $T=5$\,K.}
\label{fig4}
\end{figure} 

At fixed $\varepsilon_{j}$, phonon-induced resistivity is
periodic in inverse $B$ field with period $\Delta \varepsilon_{s\lambda}\approx 1$.
The total resistivity $r_{xx}$ roughly follows this rule, as shown in 
Fig.\,4, where $r_{xx}$ is plotted versus $B$
for fixed $\varepsilon_{j}=1$ and 1.5. Both curves show remarkable oscillations 
that have the same periodicity
but in- and out-of-phase as that of the zero-bias ($\varepsilon_{j}=0$) case and 
with much enhanced amplitudes.
These features are just those observed in the experiments.\cite{WZ-prl08}

\section{Lower mobility systems}

Magnetophonon resonance can show up in resistivity at even lower lattice temperature
as long as a finite current density is applied. In a system with lower mobility, 
the phonon contributions to resistivity can be greatly enhanced
by a modest finite current  due to the rapid rise in the electron temperature.  

\begin{figure}
\includegraphics [width=0.40\textwidth,clip=on] {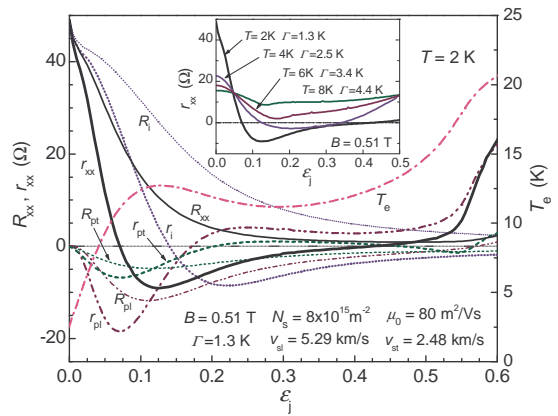}
\vspace{-0.2cm}
\caption{(Color online)  Absolute  $R_{xx}$ and differential $r_{xx}$ resistivities and their
respective contributions from impurity, LA phonons, and TA phonons  
($R_{\rm i}$ and $r_{\rm i}$, $R_{\rm pl}$ and $r_{\rm pl}$
and $R_{\rm pt}$ and $r_{\rm pt}$)
and the electron temperature $T_{\rm e}$ are shown 
versus $\varepsilon_{j}$ at a lattice temperature $T=2$\,K and a magnetic field $B=0.51$\,T 
for a GaAs-based heterosystem. The inset plots $r_{xx}$ at $T=2$, 4, 6, and 8\,K with
corresponding half-width of the Landau-level ${\it \Gamma}$.}
\label{fig5}
\end{figure}

As another example, Fig.\,4 plots the total absolute resistivity 
$R_{xx}$ and total differential resistivity $r_{xx}$ and 
their respective contributions from impurity, LA phonons, and TA phonons 
($R_{\rm i}$ and $r_{\rm i}$, $R_{\rm pl}$ and $r_{\rm pl}$
and $R_{\rm pt}$ and $r_{\rm pt}$),
and the electron temperature $T_{\rm e}$,
as functions of bias current density in terms of $\varepsilon_j$ 
at lattice temperature $T=2$\,K and magnetic field $B=0.51$\,T 
for a GaAs-based heterosystem
that has carrier sheet density $N_{\rm s}=8\times 10^{15}$m$^{-2}$ and zero-temperature
linear mobility $\mu_0=80$\,m$^2$/V\,s in the absence of magnetic field. 
The Landau-level half-width ${\it \Gamma}$ is taken to be 1.3\,K 
($2{\it \Gamma}\approx 0.25\omega_{c}$
at this magnetic field). 

Note that the phonon-induced absolute resistivities
$R_{\rm pl}$ and $R_{\rm pt}$ can become negative in this case. Although the total resistivity
$R_{xx}$ remains always positive, the growing negative  
$r_{\rm pl}$ and $r_{\rm pt}$ with 
increasing current density at the initial stage greatly accelerate the drop in the total differential 
resistivity such that $r_{xx}$ goes down and becomes negative at   
$\epsilon_j\approx 0.07$, which is much smaller than it would be without phonon contribution
($\epsilon_j\approx 0.14$), solely determined by the width of the Landau level.\cite{Lei-07-1}
As a result, the total differential resistivity is negative within a wide current range 
$0.07<\epsilon_j<0.42$ before it becomes positive with further increasing current density. 
This may provide a possible explanation for the zero-differential resistance 
state recently observed at quite small dc bias in impure samples.\cite{Bykov-prl07}

As an important parameter in the present analysis, 
the half-width $\it \Gamma$ of the Landau level, or the single particle life-time or 
 $\tau_s=1/2{\it \Gamma}$ of the electron in the magnetic field, which depends on 
all kinds of elastic and inelastic scatterings induced by electron-impurity, electron-phonon 
and electron-electron interactions in the system,
is temperature $T$ dependent. At fixed $B$ field, $\it \Gamma$ certainly increases 
with increasing temperature due to enhanced electron-phonon 
and electron-electron scatterings. With this in consideration, the sharp drop in differential
resistivity $r_{xx}$ at $T=2$\,K rapidly disappears when increasing temperature, 
as shown in the inset of Fig.\,5, which is calculated with enlarged ${\it \Gamma}$
at $T=4$, 6, and 8\,K.

\section{Summary}
In summary, we have examined magneto-phonon resonances in GaAs-based 2D systems at low temperatures.
Good agreement between theoretical prediction 
and experimental observation is obtained by using the well-defined bulk LA and TA phonon modes 
with no adjustable parameter.
We find that a finite dc excitation can not only greatly enhance the phonon resistivity 
by heating the electron gas,
but can also induce additional and prominent phonon resonance peaks, which give rise to 
remarkable resistance oscillations and negative differential resistivity in nonlinear transport. 

This work was supported by the projects of the National Science Foundation of China
and the Shanghai Municipal Commission of Science and Technology.

\end{document}